\documentclass[10pt,letterpaper,twocolumn]{article}
\usepackage[english]{babel}
\usepackage{graphicx}
\usepackage{latexsym}

\newcommand{\alt}{\mathbin{\lower 3pt\hbox
   {$\rlap{\raise 5pt\hbox{$\char'074$}}\mathchar"7218$}}}
\newcommand{\agt}{\mathbin{\lower 3pt\hbox
   {$\rlap{\raise 5pt\hbox{$\char'076$}}\mathchar"7218$}}}

\begin{document}

\setcounter{footnote}{0}
\setcounter{equation}{0}
\setcounter{figure}{0}
\setcounter{table}{0}

\title{\large\bf Strict parabolicity of the multifractal spectrum
at the Anderson transition}

\author{\small  I. M. Suslov \\
\small Kapitza Institute for Physical Problems,
\\  \small Moscow, Russia \\{} \\
\parbox{120mm}{\footnotesize  Using the well-known "algebra of
multifractality", we derive the functional equation for anomalous
dimensions $\Delta_q$, whose solution  $\Delta_q=\chi q(q-1)$
corresponds to strict parabolicity of the
multifractal spectrum. This result demonstrates clearly that a
correspondence of the nonlinear $\sigma$-models with the initial
disordered systems is not exact.
 } }

\date{}

\maketitle

Recently there has been a great interest to
multifractal properties of the wave functions, arising at the
Anderson transition point  (see a review article \cite{1}).
They are exhibited
in the anomalous scaling
$$
\langle P_q\rangle \sim L^{-D_q(q-1)} \sim
L^{-d(q-1)+\Delta_q}\,
 \eqno(1)
 $$
for the inverse participation ratios
$$
P_q=\int d^dr |\Psi({\bf r})|^{2q} \,,
 \eqno(2)
 $$
where  $\Psi({\bf r})$ is a normalized wave function of an
electron in the random potential for a finite system, having a
form of the $d$-dimensional cube with a side  $L$. In the
metallic phase  $\Psi({\bf r})$ extends along the whole system
and $|\Psi({\bf r})|^2\sim L^{-d}$ from the normalization
condition, so $P_q\sim L^{-d(q-1)}$. In the critical point
(see (1)), instead of the geometric dimension  $d$
a set of fractal dimensions $D_q$ arises, whose difference
from $d$ is determined by anomalous dimensions  $\Delta_q$.

It was noted in \cite{1} that a knowledge of anomalous
dimensions $\Delta_q$ allows to establish the behavior
of arbitrary $n$-point correlators
$$
K\{{\bf r}_{i}\} =
\left\langle |\Psi({\bf r}_1)|^{2q_1} \, |\Psi({\bf
r}_2)|^{2q_2}\,\ldots \,|\Psi({\bf r}_n)|^{2q_n} \right\rangle
\,,  \eqno(3)
$$
but the specific results were presented only for  $n=2$. It is
shown below, that consideration of the  $n>2$ case leads to
a functional equation for $\Delta_q$, whose solution corresponds
to a strictly parabolic character of the multifractal spectrum.
The analysis exploits a possibility to represent correlator
 (3) in the form of the single product of the $|{\bf r}_i-{\bf
 r}_j|$ powers, which can be justified in the small $q_i$ region
 and in fact always arises as a consequence of the matching
conditions (see the text around Eq.23 and below).

The result for $n=1$ follows from Eqs.\,1,\,2:
$$
\langle |\Psi({\bf r})|^{2q}\rangle =
L^{-d} \langle P_q\rangle \sim
L^{-dq+\Delta_q}\,.
 \eqno(4)
 $$
For $n=2$ we have, assuming a power law dependence on
$r_{12}=|{\bf r}_1-{\bf r}_2|$,
$$
\left\langle |\Psi({\bf r}_1)|^{2q_1} \, |\Psi({\bf
r}_2)|^{2q_2}\right\rangle = \,A
\left(\frac{L}{r_{12}}\right)^{\alpha}\,,
  \eqno(5)
$$
where the normalization constant $A$ and the exponent $\alpha$
can be established using the so called "algebra of
multifractality" \cite{1,2}. For $r_{12}\sim L$,
the functions
$\Psi({\bf r}_1)$ and $\Psi({\bf r}_2)$ are statistically
independent\,\footnote{\,In the localized regime, when $\xi\ll L$
($\xi$ is the localization length), the functions  $\Psi({\bf
r}_1)$ and $\Psi({\bf r}_2)$ are statistically
independent for  $r_{12}\agt \xi$, while a power law
behavior (5)
is valid for $r_{12}\alt  \xi$; both properties hold
approximately at $r_{12}\sim  \xi$.  This situation remains
unchanged, if  $\xi$ is increased to a value of
the order of $L$, i.e. at the boundary of the critical
region.  }, so the correlator (5)  reduces
to the product
$$
\left\langle |\Psi({\bf r}_1)|^{2q_1}\right\rangle \,
\left\langle |\Psi({\bf r}_2)|^{2q_2}\right\rangle \,
\sim  \,A \, \sim
$$
$$
\sim\, L^{-dq_1+\Delta_{q_1}} \cdot
L^{-dq_2+\Delta_{q_2}}\,,
  \eqno(6)
$$
which is estimated using (4).  For $r_{12}=0$, a divergency in
(5) is cut off at the atomic scale $a$,
$$
\left\langle |\Psi({\bf r}_1)|^{2q_1+2q_2} \right\rangle
\sim  \,A \left(\frac{L}{a}\right)^{\alpha}\,
\sim \, L^{-d(q_1+q_2)+\Delta_{q_1+q_2}} \,,
\eqno(7)
$$
and Eqs.\,6,\,7 lead to the results
$$
A \, \sim \, L^{-d(q_1+q_2)+\Delta_{q_1}+\Delta_{q_2}}\,, \quad
\alpha= \Delta_{q_1+q_2} - \Delta_{q_1}- \Delta_{q_2} \,, \eqno(8)
$$
in accordance  with  \cite{1,2}.
\vspace{2mm}

For the case  $n=3$ we write analogously
($r_{ij}=|{\bf r}_i-{\bf r}_j|$)
$$
\left\langle |\Psi({\bf r}_1)|^{2q_1} \,
|\Psi({\bf r}_2)|^{2q_2} \,|\Psi({\bf r}_3)|^{2q_3}
\right\rangle =
$$
$$\qquad\qquad\qquad
=\,A \, \left(\frac{L}{r_{12}}\right)^{\alpha}\,
\left(\frac{L}{r_{13}}\right)^{\beta}\,
\left(\frac{L}{r_{23}}\right)^{\gamma}\,,
  \eqno(9)
$$
and find  $A$, $\alpha$, $\beta$, $\gamma$ using the algebra of
multifractality. If all $r_{ij}\sim L$, then
$$
\left\langle |\Psi({\bf r}_1)|^{2q_1}\right\rangle \,
\left\langle |\Psi({\bf r}_2)|^{2q_2}\right\rangle \,
\left\langle |\Psi({\bf r}_3)|^{2q_3}\right\rangle \,
\sim
$$
$$
\sim  \,A \, \sim \, L^{-dq_1+\Delta_{q_1}} \cdot
L^{-dq_2+\Delta_{q_2}} \cdot L^{-dq_3+\Delta_{q_3}}
\,,
  \eqno(10)
$$
while for $r_{12}=0$, $r_{13}\sim r_{23}\sim L$ we have a result
$$
\left\langle |\Psi({\bf r}_1)|^{2q_1+2q_2}  \right\rangle \,
\left\langle|\Psi({\bf r}_3)|^{2q_3} \right\rangle
\sim
$$
$$
\sim\,A \, \left(\frac{L}{a}\right)^{\alpha}\,
\sim \, L^{-d(q_1+q_2)+\Delta_{q_1+q_2}} \cdot
 L^{-dq_3+\Delta_{q_3}}\,.
  \eqno(11)
$$
Analogous relations are valid in  cases $r_{13}=0$,
$r_{12}\sim r_{23}\sim L$ and $r_{23}=0$, $r_{12}\sim r_{13}\sim L$.
Finally, for ${\bf r}_1={\bf r}_2={\bf r}_3$ one gets
$$
\left\langle |\Psi({\bf r}_1)|^{2q_1+2q_2+2q_3} \right\rangle \,
\sim \,A \, \left(\frac{L}{a}\right)^{\alpha+\beta+\gamma}\,
\sim
$$
$$ \qquad\qquad\qquad
\sim\, L^{-d(q_1+q_2+q_3)+\Delta_{q_1+q_2+q_3}} \,,
  \eqno(12)
$$
so we have five relations for four quantities $A$, $\alpha$, $\beta$,
$\gamma$:
$$
A \, \sim \, L^{-d(q_1+q_2+q_3)+\Delta_{q_1}+\Delta_{q_2}
+\Delta_{q_3}} \,,
$$
$$
\alpha = \Delta_{q_1+q_2}-\Delta_{q_1}-\Delta_{q_2} \,,
$$
$$
\beta = \Delta_{q_1+q_3}-\Delta_{q_1}-\Delta_{q_3} \,,
\eqno(13)
$$
$$
\gamma = \Delta_{q_2+q_3}-\Delta_{q_2}-\Delta_{q_3} \,,
$$
$$
\alpha+\beta+\gamma = \Delta_{q_1+q_2+q_3}-
\Delta_{q_1}-\Delta_{q_2}-\Delta_{q_3}  \,,
$$
which cannot be satisfied for an arbitrary form of  $\Delta_q$.
For solubility of (13) a self-consistency condition should be
fulfilled
$$
\Delta_{q_1+q_2+q_3} = \Delta_{q_1+q_2}
+\Delta_{q_1+q_3}+\Delta_{q_2+q_3}-
\Delta_{q_1}-\Delta_{q_2}-\Delta_{q_3}  \,,
\eqno(14)
$$
which is a functional equation for  $\Delta_q$. It is easy to
verify that Eq.14 is satisfied for the spectrum
$\Delta_q=aq^2+bq$, and in fact it is the only possible form.
Indeed, setting $q_1=q$, $q_2=q_3=\delta$, one has
$$
\Delta_{q+2\delta} = 2 \Delta_{q+\delta}-\Delta_{q}
+\Delta_{2\delta}-2\Delta_{\delta}  \,,
\eqno(15)
$$
and expansion to the second order in  $\delta$ gives
$$
\Delta''_{q} =  \Delta''_{0} \,,
\eqno(16)
$$
where we have used the condition  $\Delta_0=0$ derived from
Eqs.1,\,2.  Since  $\Delta''_{0}$ is simply a constant, one can
integrate (16) and  obtain an arbitrary quadratic polynomial in
$q$, which reduces to a form $\Delta_q=aq^2+bq$, if the equality
$\Delta_0=0$ is exploited. In the absence of singularities on the
$q$-axis, one can use another relation $\Delta_1=0$
obtained from Eqs.1,\,2 and arrive to the finite
form\,\footnote{\,In fact, for validity of (17) one needs the
absence of singular points in the interval $(0,1)$, which is
confirmed by numerical experiments for dimensions $d=2,\,3,\,4$.
In the general case, one should use the form $\Delta_q=aq^2+bq$
in each interval of regularity, so dependence $\Delta_{q}$ may
consist of several parabolic or linear pieces.  There are
indications that such variant is realized in high dimensions. }
$$
\Delta_{q} =  \chi q(q-1) \,, \qquad \chi>0 \,.
\eqno(17)
$$
The positiveness of $\chi$ follows from inequality  $\tau''_q\le
0$, where $\tau_q=D_q(q-1)$ \cite{1}.  \vspace{2mm}

In the case of the general $n$-point correlator we accept
$$
\left\langle |\Psi({\bf r}_1)|^{2q_1} \, |\Psi({\bf
r}_2)|^{2q_2}\,\ldots \,|\Psi({\bf r}_n)|^{2q_n} \right\rangle
=\,A\, \prod\limits_{i<j}
\left(\frac{L}{r_{ij}}\right)^{\alpha_{ij}}\,
\eqno(18)
$$
and obtain analogously to the preceding
$$
A \, \sim \, L^{-d(q_1+q_2+\ldots+q_n)+\Delta_{q_1}+\Delta_{q_2}
+\ldots+\Delta_{q_n}} \,,
$$
$$
\alpha_{ij} = \Delta_{q_i+q_j}-\Delta_{q_i}-\Delta_{q_j} \,.
\eqno(19)
$$
Rewriting the product (18) in the form clarifying its
dependence  on  $r_{i,n}$ and $r_{i,n-1}$
$$
 \prod\limits_{i=1}^{n-1}   \prod\limits_{j=i+1}^n
\left(\frac{L}{r_{ij}}\right)^{\alpha_{ij}}\, =\,
 \prod\limits_{i=1}^{n-1}
\left(\frac{L}{r_{i,n}}\right)^{\alpha_{i,n}} \cdot
$$
$$
\cdot
 \prod\limits_{i=1}^{n-2}
\left(\frac{L}{r_{i,n-1}}\right)^{\alpha_{i,n-1}}
 \prod\limits_{i=1}^{n-3}   \prod\limits_{j=i+1}^{n-2}
\left(\frac{L}{r_{ij}}\right)^{\alpha_{ij}}
\eqno(20)
$$
and setting  ${\bf r}_{n-1}={\bf r}_n$, one has
$$
\left\langle |\Psi({\bf r}_1)|^{2q_1} \, |\Psi({\bf
r}_2)|^{2q_2}\,\ldots \,|\Psi({\bf r}_{n-1})|^{2q_{n-1}+2q_n}
\right\rangle  \sim
$$
$$
\qquad\sim\,A\, \left(\frac{L}{a}\right)^{\alpha_{n-1,n}}
 \prod\limits_{i=1}^{n-2}
\left(\frac{L}{r_{i,n-1}}\right)^{\alpha_{i,n-1}+\alpha_{i,n}}
\cdot
$$
$$\qquad\qquad\qquad\cdot
 \prod\limits_{i=1}^{n-3}   \prod\limits_{j=i+1}^{n-2}
\left(\frac{L}{r_{ij}}\right)^{\alpha_{ij}}  \,,
\eqno(21)
$$
which should be consistent with the result for the $(n-1)$-point
correlator, obtained from (18) by replacements
 $n\to n-1$ and $q_{n-1} \to q_{n-1}+q_{n}$.
A self-consistency condition reduces to the equality
$$
\Delta_{q_i+q_{n-1}+q_n} = \Delta_{q_i+q_{n-1}}
+\Delta_{q_i+q_n}+\Delta_{q_{n-1}+q_n}-
$$
$$
-\Delta_{q_i}-\Delta_{q_{n-1}}-\Delta_{q_n}  \,,
\eqno(22)
$$
which is analogous to (14) and  satisfied for the parabolic
spectrum. We see that a functional form (17) provides
self-consistency of results  (18),\,(19) for arbitrary
$n$-point correlators.

Above we have accepted that  correlator (3) is
determined by a single product of the $r_{ij}$ powers.
Generally, the right hand side of (18) may contain less
singular terms determined by exponents  $\tilde\alpha_{ij}$,
whose sum is less than a sum of $\alpha_{ij}$. If certain
$\tilde\alpha_{ij}$ are greater than $\alpha_{ij}$, then
the given analysis becomes invalid. The absence of such terms
can be established for sufficiently small  $q_i$. Indeed,
expansion of (18) over $q_i$ with $\Delta_q=aq^2+bq$
shows that for validity of (19) one should set
$$
\left\langle \ln|\Psi({\bf r}_i)|^2\right\rangle = (b-d)\ln L\,,
$$
$$
\left\langle \ln^2|\Psi({\bf r}_i)|^2 \right\rangle =
(b-d)^2\ln^2 L + 2a\ln{L}\,,
$$
$$
\left\langle \ln|\Psi({\bf r}_i)|^2 \ln|\Psi({\bf r}_j)|^2
\right\rangle = (b-d)^2\ln^2 L + 2a\ln(L/r_{ij})\,.
\eqno(23)
$$
These relations are valid for $n=2$, if the power law dependence
is accepted in (5), and then they automatically
hold for arbitrary $n$, justifying representation (18).
If additional terms are present in (18), then relations analogous
to (23) can be fulfilled only in the presence of certain
relations between the exponents  $\alpha_{ij}$ and
$\tilde\alpha_{ij}$.
It is clear from Wilson's many-parameter renormalization
group that the main scaling and corrections to
it\,\footnote{\,The arbitrary choice of $q_{i}$ allows
to neglect the exceptional situations when the sum of
$\alpha_{ij}$ is equal to the sum of $\tilde\alpha_{ij}$,
and separate the main contribution from corrections to it.  }
originate from different sources and appear to be independent;
so existence of strict relations between  $\alpha_{ij}$
and $\tilde\alpha_{ij}$ looks
improbable\,\footnote{\,Such relations are possible in
conformal theories, which possess deep internal symmetry.
However, relation (9) for $n=3$ is exact in conformal theories
 \cite{100}. }.
Thereby, for sufficiently small $q_i$ there are
no additional terms in (18), so the spectrum is strictly
parabolic in a certain vicinity of $q=0$ and can be
analytically continued to any interval, not containing singular
points.\,\footnote{\,For finite $L$, analiticity of $P_q$ and
$\Delta_q$ follows from definition (2) according to the theorem
on analyticity of integrals depending on a parameter (see,
for example the book \cite{101}). In the limit $L\to\infty$ there
is a possibility of isolated singular points due to the
reasons analogous to the Stokes phenomenon (a change of
topology for the steepest descent trajectories ); such
singularities are discussed in Sec.\,II.C.7 of the paper [1].}.
The latter reservation is essential, because existence of
singular points looks rather probable  (see below).

The structure of correlators used in the paper can
be justified using the well-known operator product
expansion [3]
$$
A_l({\bf r}_1)\, A_m({\bf r}_2)\, = \,\sum\limits_k
C^k_{lm}({\bf r}_1\!-\!{\bf r}_2) \,A_k({\bf r}_2) \,,
\eqno(24)
$$
which allows to produce successive diminishing of the
order of the correlator
$$
\left\langle A_1({\bf r}_1) A_2({\bf r}_2)\ldots A_n({\bf r}_n)
\right\rangle
\eqno(25)
$$
and represent it as the sum of products of the coefficient
functions  $C^k_{lm}({\bf r}_i\!-\!{\bf r}_j)$. The latter
naturally have a power-law behavior at the critical point,
leading to representation of correlators as sums of products
composed from the $r_{ij}$ powers. Such representation is not
unique, because a pair of operators in (24) can be chosen in
different ways, and the result depends on succession the
operators are chosen in the course of reducing of the
correlator. It gives the functional relations between
$C^k_{lm}({\bf r}_i\!-\!{\bf r}_j)$, which allow to
transfer from one representation to another. One can trace
on example of conformal theories  [3],  how to
obtain the representation containing the product of all
$r_{ij}$ in the leading term; such representation is
implied in the present paper.

Existence of such representation is not related with
specificity of the conformal theory. Indeed, let consider the
case $n=3$ as an example. Suggesting that ${\bf r}_i$ is
close to  ${\bf r}_j$, we apply the operator product expansion
to the pair of operators $(i,j)$ and retain the leading terms in
$r_{ij}$; then the following results are obtained for the
correlator $K\{{\bf r}_{i}\}$:
$$
K\{{\bf r}_{i}\} = (r_{12})^{-\alpha} \, f_1(r_{13},r_{23})\,,
\qquad r_{12} \ll r_{13} \approx r_{23} \,,
$$
$$
K\{{\bf r}_{i}\} = (r_{13})^{-\beta} \, f_2(r_{12},r_{23})\,,
\qquad r_{13} \ll r_{12} \approx r_{23}  \,,
\eqno(26)
$$
$$
K\{{\bf r}_{i}\} = (r_{23})^{-\gamma} \, f_3(r_{12},r_{13})\,,
\qquad r_{23} \ll r_{12} \approx r_{13}   \,.
$$
The correct form of functions  $f_i$ cannot be
established, because the corresponding two arguments are
indistinguishable in this limit. For coinciding arguments
these functions have a power law behavior, which is
partially related with the first, and partially with the
second argument:
$$
K\{r_{ij}\} = (r_{12})^{-\alpha} \, (r_{13})^{-\beta'}\,
(r_{23})^{-\gamma'}\,, \qquad r_{12} \ll r_{13} \approx r_{23}
\,,
$$
$$
K\{r_{ij}\} = (r_{13})^{-\beta} \,
(r_{12})^{-\alpha'}\,(r_{23})^{-\gamma''}
\,, \qquad r_{13} \ll r_{12} \approx r_{23}
\,,
\eqno(27)
$$
$$ K\{r_{ij}\} = (r_{23})^{-\gamma} \,
(r_{12})^{-\alpha''}\,(r_{13})^{-\beta''}\,,
\qquad r_{23} \ll r_{12} \approx r_{13}
\,.
$$
If all three configurations are different, then the operator
product expansion contains three essentially different terms
with the same sum of exponents  (it is clear from consistency
of expressions for coinciding  $r_{ij}$). Such degeneracy is
natural in the case  $q_1=q_2=q_3$; for unequal exponents it
arises inevitably in the course of symmetrization over $r_{ij}$.
However, in the present paper  (in opposite to  [8])
we consider configurations $\{q_i\}$ of the general position;
then such degeneracy is not supported by symmetry and looks
completely improbable. Hence, we deal with one and the same
configuration in  (27), i.e.
$$
K\{{\bf r}_{i}\} \sim (r_{12})^{-\alpha} \, (r_{13})^{-\beta} \,
(r_{23})^{-\gamma}
\eqno(28)
$$
in correspondence with Eq.9. The exponents $\alpha$, $\beta$,
$\gamma$ are inevitably determined by the second formula in
(19); setting all $r_{ij}\sim a$, or  $r_{ij}\sim L$, in Eq.27
gives two extra relations, so all equations (13) are reproduced.
It is clear from this reasoning that a strict
validity of (9) is not necessary for our analysis, because this
relation  arises effectively  due to matching conditions
for three formulas (26).

The parabolic spectrum  (17) corresponds to the logarithmically
normal distribution for the amplitudes $|\Psi({\bf r})|^2$
\cite{3}. If the latter distribution is accepted axiomatically,
then  (17) is valid for arbitrary  $q$. The
statement of  \cite{1} on impossibility of such a situation refers
to the lattice models, where inequality $|\Psi({\bf r})|^2\le 1$
holds due to discreteness of the coordinate ${\bf r}$
(the equality $|\Psi({\bf r}_0)|^2= 1$ corresponds to
localization at the single site  ${\bf r}_0$). This inequality
leads to restriction $\alpha\ge 0$ for the definitional domain
of the singularity spectrum  $f(\alpha)$ and impossibility of
the decreasing behavior for  $\tau_q=D_q(q-1)$;\,\footnote{\,The
function $f(\alpha)$ is related with $\tau_q$  via the Legendre
transformation $\tau_q=q\alpha-f(\alpha)$, $q=f'(\alpha)$.
In particular, $f(\alpha)=d-(\alpha-\alpha_0)^2/4(\alpha_0-d)$
with $\alpha_0=d+\chi$ for the spectrum (17).}
 as a result,
dependence $\tau_q$ saturates by a constant $\tau_{q_c}$ for
$q>q_c$, where $q_c$ is a certain singular point. These
restrictions are inessential for continuous models, where the
parabolic spectrum is possible for arbitrary $q$.

Nevertheless, one cannot exclude the existence of singular
points, since the algebra of multifractality is certainly
violated for large positive  $q_i$. Indeed, setting  $q_2=1$
in (5) and integrating over  ${\bf r}_2$, one can easily test
that results (8) are valid only for  $\alpha\le d$, which
corresponds to  $q_1\le d/2\chi$ for the spectrum (17). Violation
of algebra is a consequence of quick decreasing of functions
$|\Psi({\bf r}_i)|^{2q_i}$  at large distances from their
"centers", so they become statistically independent at
a scale of $r_{ij}$ lesser than  $L$.

In the approach based on the use of nonlinear $\sigma$-models
\cite{4}, parabolicity of the spectrum takes place for the
spatial dimension $d=2+\epsilon$ in the lowest orders in
$\epsilon$  \cite{1,4}, but is violated on the four-loop level.
This situation is not unexpected: derivation of $\sigma$-models
is justified only for small  $\epsilon$, and the question on
their exact correspondence with the initial disordered systems
always remained open. In particular, strong doubts arose in
relation with the upper critical dimension \cite{5}.  The paper
\cite{6} suggests explanation why deficiency of $\sigma$-models
for the orthogonal ensemble arises just on the four-loop
level\,\footnote{\,For the unitary ensemble, the paper \cite{6}
gives the simple and completely rigorous proof of the
$\sigma$-model deficiency based on inequality for $\Delta_q$.  }.
In the "minimal" $\sigma$-model used by Wegner, one is restricted
by the lower (second) powers of gradients, which corresponds to
neglecting the spatial dispersion of the diffusion coefficient
$D(\omega,q)$. In the first three orders in $\epsilon$ this
approximation is self-consistent, while self-consistency
fails on the four-loop level. As a result, one should add
the terms
with higher gradients which leads to instability of the
renormalization group due the "gradient catastrophe"
\cite{7}. To remove instability one should include the additional
counter-terms, which leads to essential modification of the
$\sigma$-model Lagrangian and inevitable revisiting of all
four-loop contributions. The latter may eliminate a discrepancy
with self-consistent theory by Vollhardt and W$\ddot o$lfle
\cite{8}, or its refined version \cite{8a}.

A surprising accuracy of Wegner's one-loop result  \cite{4}
(corresponding to  (17) with $\chi=\epsilon$) in application to
the $d=3$ and $d=4$ cases was
reported in a lot of numerical
experiments \cite{9,10,11,12,13}, though detectable
deviations were also declared (Figs.1--3). For example, a
position of the maximum for the singularity
spectrum $f(\alpha)$
(which is $\alpha_0=d+\epsilon$ in the one-loop approximation
\cite{1,4}) was estimated as  $\alpha_0=4.03\pm 0.05$ \cite{9},
$\alpha_0=4.048\pm 0.003$ \cite{12}\,\footnote{\,Estimation of
errors in the paper \cite{12} arouses serious doubts
(see Footnote 11 in \cite{6}). }
 for $d=3$  and
$\alpha_0=6.5\pm 0.2$ \cite{9}  for $d=4$.
A value $\alpha_0$ corresponds to the maximum of the
distribution function for  $\ln |\Psi({\bf r})|^2$, where
numerical data are the most reliable, while their accuracy becomes
worse near the tails of distribution (Figs.\,1,\,2). In whole,
the parabolic form of the spectrum is confirmed on the level of
10\%.  As demonstrated in Sec.5 of the paper  \cite{6},
convergence of  correlators (3) to the
thermodynamic limit is extremely slow, and a systematic error for
fractal dimensions can reach tens of percents. Therefore, the
observed deviations from parabolicity (Fig.3) are surely within
expectations.

\begin{figure}
\centerline{\includegraphics[width=3.1 in]{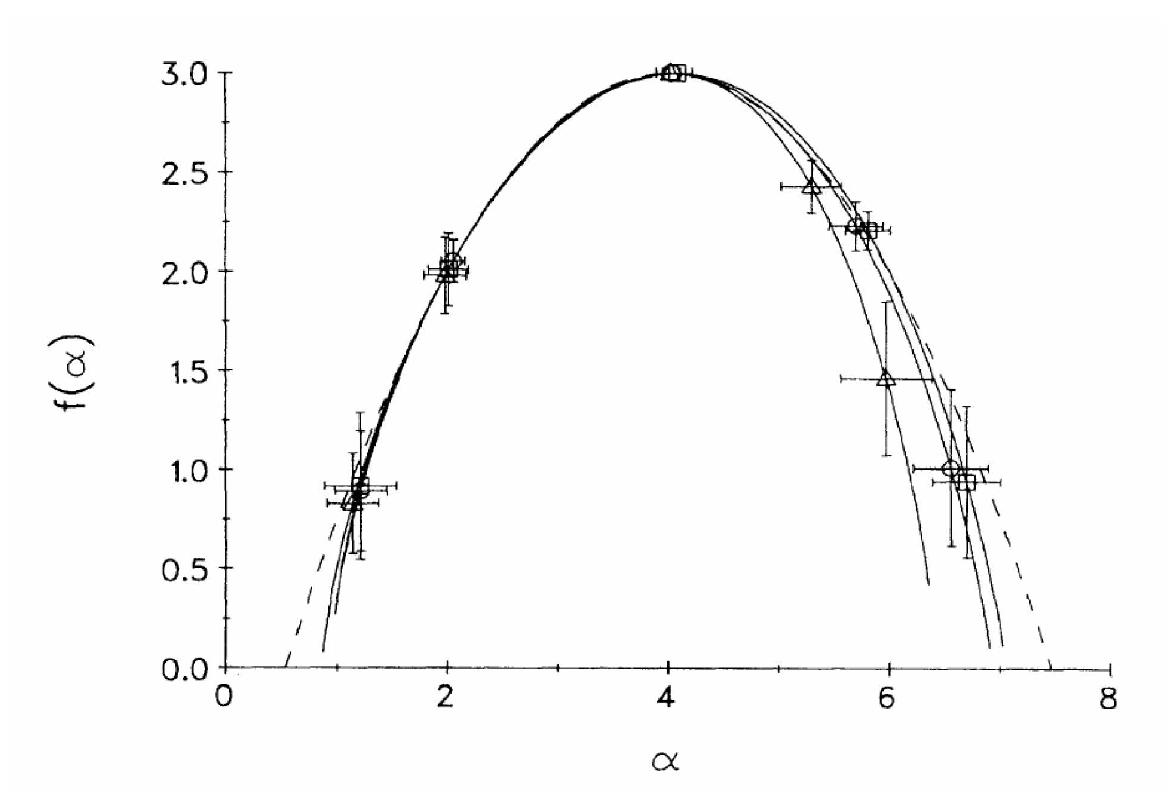}} \caption{
Singularity spectrum $f(\alpha)$  for the Anderson model
with box ($\Large\circ$), Gaussian ($\Box$), and binary
($\small\triangle$) distribution \cite{10}.  The dashed line
shows the one-loop Wegner result.  }
\label{fig1}
\end{figure}

\begin{figure}
\centerline{\includegraphics[width=2.8 in]{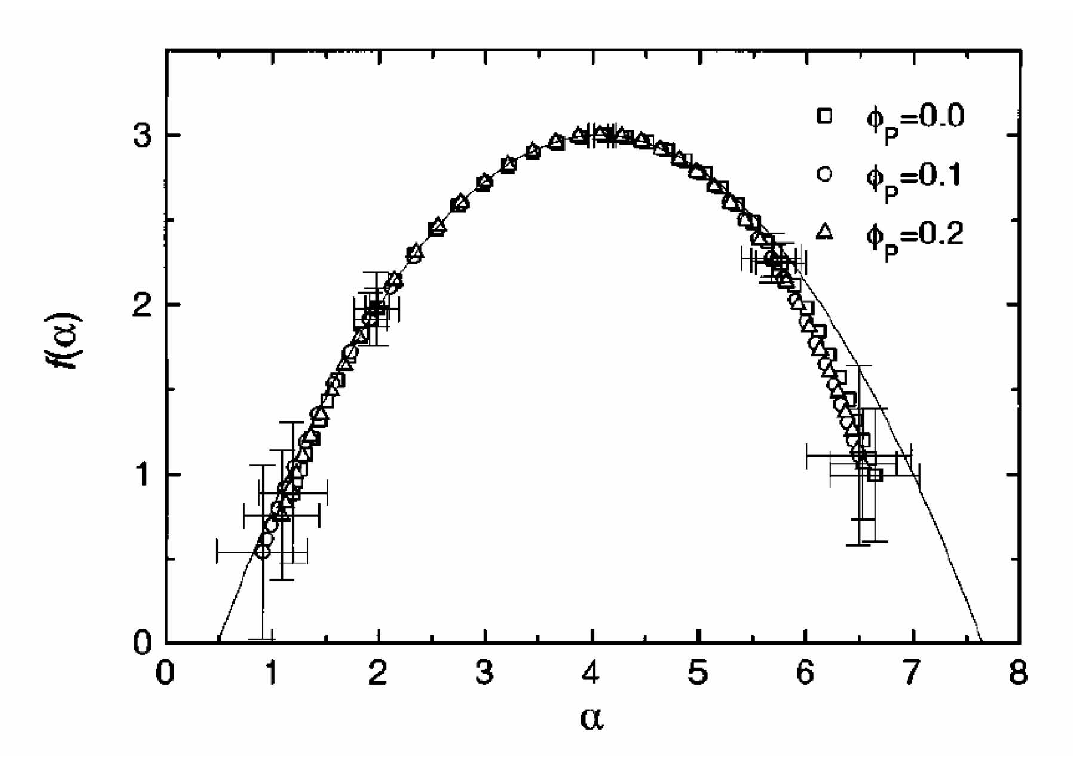}} \caption{
Singularity spectrum $f(\alpha)$ in the absence of the magnetic
field ($\phi_P=0.0$) and for two different magnitude of field
\cite{11}.
The solid line corresponds to the parabolic spectrum with
$\alpha_0=4.1$. }
\label{fig2} \end{figure}

\begin{figure}
\centerline{\includegraphics[width=2.8 in]{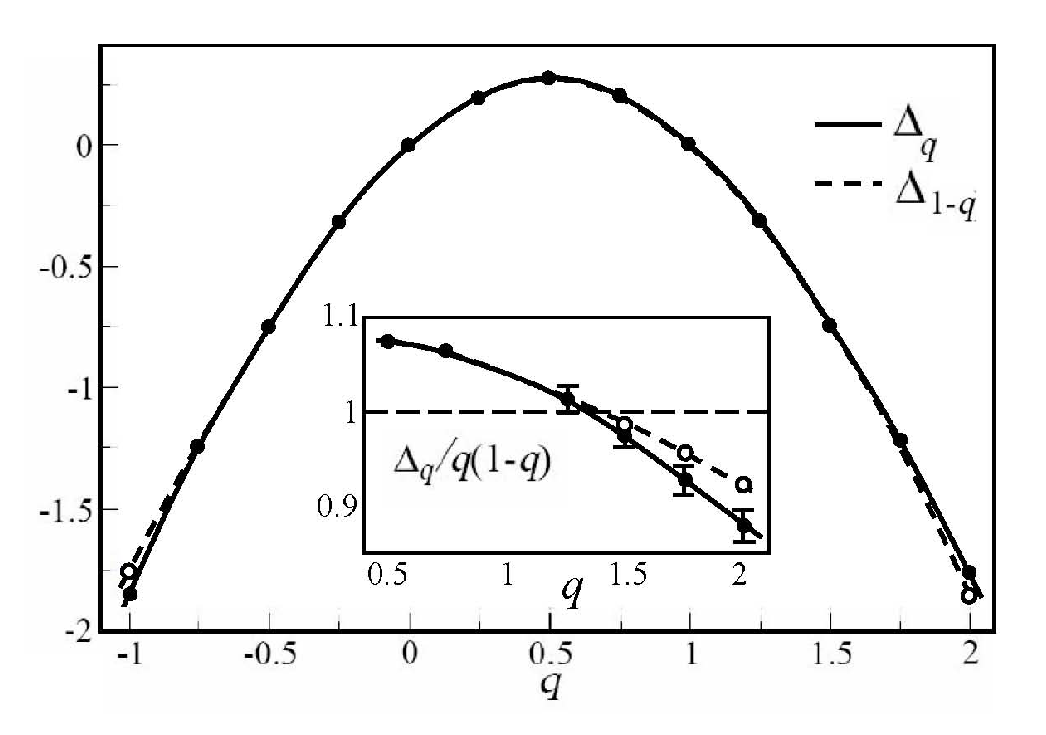}} \caption{
Multifractal exponents $\Delta_q$ (defined with the
opposite sign), obtained from finite size scaling
\cite{12}. The inset show the reduced exponents
$\Delta_q/q(1-q)$; the horizontal dotted line corresponds to
the one-loop Wegner result. }
\label{fig3} \end{figure}

In the regime of the integer quantum Hall effect, the spectrum
is parabolic on the level of $10^{-3}$ and there are
theoretical arguments in favour of exact parabolicity
\cite{14,15,16} based on the relation with the conformal field
theory. Nevertheless, small significant deviations were
reported in \cite{13}. Such tiny deviations are unnatural, since
there are no small parameters in the system. In our opinion,
these deviations are related with slow convergence to the
thermodynamic limit, though analysis of \cite{6} is not
directly applicable here\,\footnote{\,The same point of view
was expressed by M.\,R.\,Zirnbauer
(private communication), since
deviations from parabolicity detected in \cite{17} were found to
be  related with  finite-size effects.  }.

The above considerations are not applicable to the so called
PRBM model \cite{1}, where strong deviations from parabolicity
are obtained analytically and confirmed by numerical
simulations. This model corresponds to disordered systems with
power-law correlations of a random potential. It is clear from
the example of ferromagnets with long range interaction, that
such models possess a lot of pathological properties, which are
revealed in different aspects and demand a special analysis for
detection. In the present case, there is unclear question on the
possibility to consider wave functions as statistically
independent at some scale, if the random potential is strongly
correlated in the whole system. This question should be answered
to establish validity  of the "algebra of multifractality"
(see Footnote 1).

In conclusion, the use of the algebra of multifractality
\cite{1,2} leads to the parabolic spectrum of anomalous
dimensions and clearly demonstrates that a
correspondence of the $\sigma$-models with the initial
disordered systems is not exact.



\end{document}